\documentclass[dvipdfmx]{ptephy_v1}

\preprintnumber{UTHEP-760, UTCCS-P-139} 

\begin{document}

\title{Metal-insulator transition in (2+1)-dimensional Hubbard model with tensor renormalization group}


 \author{Shinichiro Akiyama}
  \affil{Graduate School of Pure and Applied Sciences, University of Tsukuba, Tsukuba, Ibaraki 305-8571, Japan \email{akiyama@het.ph.tsukuba.ac.jp}}

  \author{Yoshinobu Kuramashi}
  \affil{Center for Computational Sciences, University of Tsukuba, Tsukuba, Ibaraki 305-8577, Japan \email{kuramasi@het.ph.tsukuba.ac.jp}}

  \author[3]{Takumi Yamashita\footnote{\it  Present address: 28-20 Kojogaoka, Otsu, Shiga 520-0821 Japan}}
  \affil{Center for Artificial Intelligence Research, University of Tsukuba, Tsukuba, Ibaraki 305-8577, Japan \email{yamashita.takumi.fn@alumni.tsukuba.ac.jp}}


\begin{abstract}%
  We investigate the doping-driven metal-insulator transition of the (2+1)-dimensional Hubbard model in the path-integral formalism with the tensor renormalization group method. We calculate the electron density $\langle n\rangle$ as a function of the chemical potential $\mu$ choosing three values of the Coulomb potential with $U=80$, $8$, and $2$ as representative cases of the strong, intermediate, and weak couplings. We have determined the critical chemical potential at each $U$, where the Hubbard model undergoes the metal-insulator transition from the half-filling plateau with $\langle n\rangle=1$ to the metallic state with $\langle n\rangle > 1$. 
Our results indicate that the model exhibits the metal-insulator transition over the vast region of the finite coupling $U$.
\end{abstract}

\subjectindex{I44}

\maketitle

\section{Introduction}
\label{sec:intro}

The Hubbard model, which is a simple theoretical model to describe electron systems with repulsive Coulomb interactions, is expected to have rich phase structures so that it has been attracting the interest of not only condensed matter physicists but also particle physicists. It has been widely known that the Hubbard model has a similar path-integral form to the Nambu--Jona-Lasinio (NJL) model~\cite{Nambu:1961tp,Nambu:1961fr}, which is a low energy effective theory in Quantum Chromodynamics (QCD): Both consist of a hopping term and a four-fermi interaction term. Their similarity, unfortunately, leads to sharing the so-called sign problem, which is a notorious difficulty in the numerical analyses based on the Monte Carlo approach. 

Recently the authors have successfully applied the tensor renormalization group (TRG) method\footnote{In this paper the TRG method or the TRG approach refers to not only the original numerical algorithm proposed by Levin and Nave \cite{Levin:2006jai} but also its extensions \cite{PhysRevB.86.045139,Shimizu:2014uva,Sakai:2017jwp,Adachi:2019paf,Kadoh:2019kqk,Akiyama:2020soe,adachi2020bondweighted,Kadoh:2021fri}.} to investigate the phase transition of the four-dimensional ($4d$) NJL model at high density and very low temperature~\cite{Akiyama:2020soe}. This work was followed by the application of the TRG method to analyze the metal-insulator transition of the $(1+1)d$ Hubbard model by calculating the electron density $\langle n\rangle$ as a function of the chemical potential $\mu$~\cite{Akiyama:2021xxr}. Our results for the critical chemical potential $\mu_{\rm c}$ and the critical exponent $\nu$ are consistent with an exact solution based on the Bethe ansatz~\cite{PhysRevLett.20.1445,LIEB20031}.

In this paper, we apply the TRG method to investigate the doping-driven metal-insulator transition in the $(2+1)d$ Hubbard model. \footnote{The model has also been investigated by the tensor network method based on the Hamiltonian formalism, like a fermionic PEPS, which is also free from the sign problem. For a recent study, see Ref.~\cite{Schneider:2021hqj}, for example.} The TRG method, which was originally proposed to study two-dimensional (2$d$) classical spin systems~\cite{Levin:2006jai}, has been developed to study wide varieties of fermionic models in particle physics~\cite{Shimizu:2014uva,Shimizu:2014fsa,Shimizu:2017onf,Takeda:2014vwa,Sakai:2017jwp,Yoshimura:2017jpk,Kadoh:2018hqq,Kadoh:2019ube,Kuramashi:2019cgs,Akiyama:2020ntf,Akiyama:2020soe,PhysRevD.101.094509}. It is also confirmed that the TRG method does not suffer from the sign problem by studying various quantum field theories~\cite{Shimizu:2014uva,Shimizu:2014fsa,Shimizu:2017onf,Takeda:2014vwa,Kawauchi:2017dnj,Kadoh:2018hqq,Kadoh:2019ube,Kuramashi:2019cgs,Akiyama:2020ntf,Akiyama:2020soe,Akiyama:2021xxr,Bloch:2021mjw,Nakayama:2021iyp}. We calculate the electron density $\langle n\rangle$ as a function of $\mu$ with three choices of $U=80$, $8$ and $2$.
The $\mu$ dependence of $\langle n\rangle$ allows us to determine the critical chemical potential $\mu_{\rm c}$ at the doping-driven metal-insulator transition from the half-filling plateau with $\langle n\rangle=1$ to the metallic state with $\langle n\rangle>1$. Our results at $U=80$, 8 and 2 show that $\vert \mu_{\rm c}- U/2\vert$ monotonically diminishes as $U$ decreases and seems to converge on $\vert \mu_{\rm c}- U/2\vert=0$ at $U=0$. This indicates the possibility that the model exhibits the metal-insulator transition over the wide region of the finite coupling.

This paper is organized as follows. In Sec.~\ref{sec:method} we define the Hubbard model in the path-integral formalism and give a brief description of the numerical algorithm. In Sec.~\ref{sec:results} we present the $\mu$ dependence of the electron density and determine the critical chemical potential $\mu_{\rm c}$ at the doping-driven metal-insulator transition.  Section~\ref{sec:summary} is devoted to summary and outlook.

\section{Formulation and numerical algorithm}
\label{sec:method}

\subsection{(2+1)-dimensional Hubbard model in the path-integral formalism}
\label{subsec:action}

We consider the partition function of the Hubbard model in the path-integral formalism on an anisotropic lattice with the physical volume $V=L_x\times L_y\times \beta$, whose spatial extension is defined as $L_\sigma=aN_{\sigma}~(\sigma=x, y)$ with $a$ the spatial lattice spacing. $\beta$ denotes the inverse temperature, which is divided as $\beta=1/T=\epsilon N_\tau$.
Following Ref.~\cite{Akiyama:2021xxr}, the path-integral expression of the partition function is given by
\begin{align}
	Z=\int\left(\prod_{n\in\Lambda_{2+1}}\prod_{s=\uparrow,\downarrow}{\rm d}\bar{\psi}_{s}(n){\rm d}\psi_{s}(n)\right){\rm e}^{-S},
	\label{eq:Z}
\end{align}
where $n=(n_x,n_y,n_{\tau})\in\Lambda_{2+1}(\subset\mathbb{Z}^3)$ specifies a position on the lattice $|\Lambda_{2+1}|=N_x\times N_y\times N_\tau$. Since the Hubbard model describes the spin-1/2 fermions, they are labeled by $s=\uparrow,\downarrow$, corresponding to the spin-up and spin-down, respectively. Introducing the notation,
\begin{align}
	\psi(n)=\left(
	\begin{array}{c}
	 	\psi_\uparrow(n)\\ \psi_\downarrow(n) 
	\end{array}
	\right),
	~\bar{\psi}(n)=\left(\bar{\psi}_\uparrow(n),\bar{\psi}_\downarrow(n)\right),
\end{align}
the action $S$ is defined as
\begin{align}
\label{eq:action}
   	S&=\sum_{n_\tau,n_x,n_y}\epsilon a^2\left\{\bar{\psi}(n)\left(\frac{\psi(n+{\hat \tau})-\psi(n)}{\epsilon}\right)\right.\nonumber\\
   	&\left.-t\sum_{\sigma=x,y}\left(\bar{\psi}(n+{\hat\sigma})\psi(n)+\bar{\psi}(n)\psi(n+{\hat\sigma})\right)+\frac{U}{2}\left(\bar{\psi}(n)\psi(n)\right)^2-\mu\bar{\psi}(n)\psi(n)\right\}.
\end{align}
The kinetic terms in the spatial directions contain the hopping parameter $t$. The four-fermi interaction term represents the Coulomb repulsion of electrons at the same lattice site. The chemical potential is denoted by the parameter $\mu$. Note that the half-filling is realized at $\mu=U/2$ in the current definition. We assume the periodic boundary condition in the spatial direction, $\psi(N_{x}+1,n_y,n_\tau)=\psi(1,n_y,n_\tau)$ and $\psi(n_x,N_{y}+1,n_\tau)=\psi(n_x,1,n_\tau)$, while the anti-periodic one in the temporal direction, $\psi(n_x,n_y,N_\tau+1)=-\psi(n_x,n_y,1)$. In the following discussion, we always set $a=1$.

\subsection{Numerical algorithm}
\label{subsec:algorithm}

Based on Ref.~\cite{Akiyama:2020sfo}, the tensor network representation of Eq.~\eqref{eq:Z} is immediately obtained. Set $d=2$ in the Appendix of Ref.~\cite{Akiyama:2021xxr} and one can find out the Grassmann tensor which generates the Grassmann tensor network of Eq.~\eqref{eq:Z}. The resulting Grassmann tensor $\mathcal{T}_{\Psi_{x}\Psi_{y}\Psi_{\tau}\bar{\Psi}_{\tau}\bar{\Psi}_{y}\bar{\Psi}_{x}}$ is of rank 6 and we evaluate $\mathrm{gTr}[\prod_{n}\mathcal{T}]$ with the anisotropic TRG (ATRG) algorithm \cite{Adachi:2019paf} whose extension to the Grassmann integrals, referred as the Grassmann ATRG (GATRG), is given in Ref.~\cite{Akiyama:2020soe}. We also follow the coarse-graining procedure employed in the previous study of $(1+1)d$ Hubbard model \cite{Akiyama:2021xxr}. Firstly, we carry out $m_{\tau}$ times of renormalization along with the temporal direction, which can be seen as the imaginary time evolution of the local Grassmann tensor. Secondly, $3d$ ATRG procedure is applied as the spacetime coarse-graining. As in the case of $(1+1)d$ Hubbard model, we have found that the optimal $m_{\tau}$ satisfies the condition $\epsilon 2^{m_{\tau}}\sim O(10^{-1})$ in the sense of preserved tensor norm. 

\section{Numerical results} 
\label{sec:results}
 
\subsection{Algorithmic-parameter dependence}

\begin{figure}
	\centering
	\includegraphics[keepaspectratio,scale=0.5]{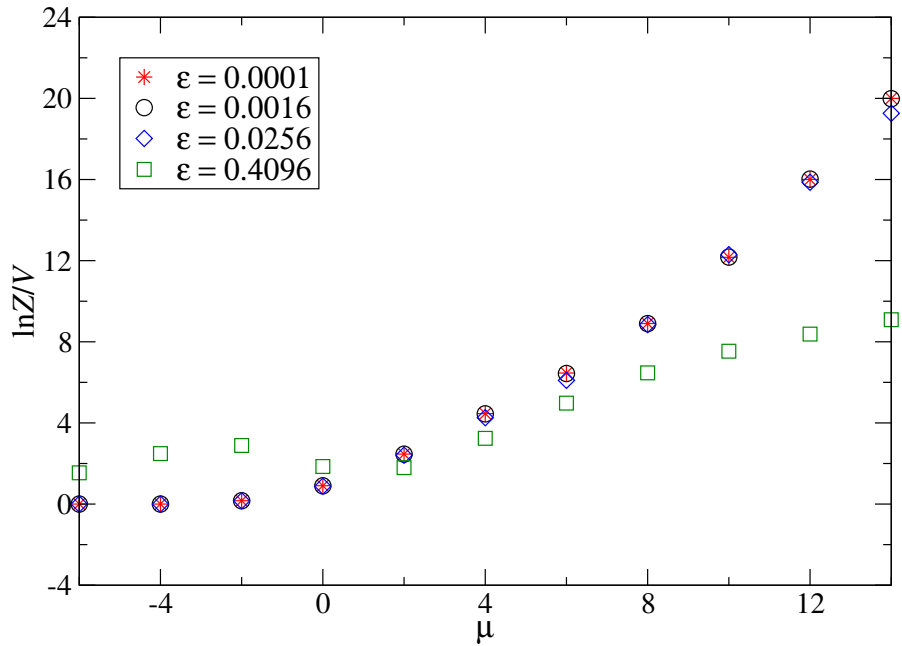}
 	\caption{Thermodynamic potential at $U=8$ on $V=4096^2\times 1677.7216$ lattice. $\beta$ is divided with $\epsilon=2^{12}\times10^{-4}=0.4096$, $2^{8}\times10^{-4}=0.0256$, $2^{4}\times10^{-4}=0.0016$ and $10^{-4}=0.0001$. The bond dimension is set to be $D=80$.}
  	\label{fig:lnZ_delbeta}
\end{figure}

The partition function of Eq.~\eqref{eq:Z} is evaluated using the numerical algorithm explained above on lattices with the physical volume $V=L_x\times L_y\times \beta=N_x\times N_y\times \epsilon N_\tau$ with $(N_{x}, N_{y}, N_\tau)=(2^m,2^m,2^{m_{\tau}+m})~(m, m_{\tau}\in\mathbb{N})$.
We employ $U=80$, 8 and 2 for the four-fermi coupling with  $t=1$ for the hopping parameter. In Fig.~\ref{fig:lnZ_delbeta} we plot the $\mu$ dependence of the thermodynamic potential $\ln Z/V$ at $U=8$ on $V=4096^2\times 1677.7216$ with the bond dimension $D=80$ in the GATRG algorithm choosing $\epsilon=2^{12}\times10^{-4},2^{8}\times10^{-4},2^{4}\times10^{-4},10^{-4}$. For each value of $\epsilon$, $m_{\tau}$ is chosen via the condition $\epsilon 2^{m_{\tau}}=2^{12}\times10^{-4}=O(10^{-1})$ following Ref.~\cite{Akiyama:2021xxr}. We find clear discretization effects for the coarsest case of $\epsilon=2^{12}\times10^{-4}$. On the other hand, the results with $\epsilon=2^{4}\times10^{-4}$ and $10^{-4}$ show good consistency. This means that the discretization effects with $\epsilon=10^{-4}$ are negligible. We employ $\epsilon=10^{-4}$ in the following calculations.

\begin{figure}
	\centering
	\includegraphics[keepaspectratio,scale=0.5]{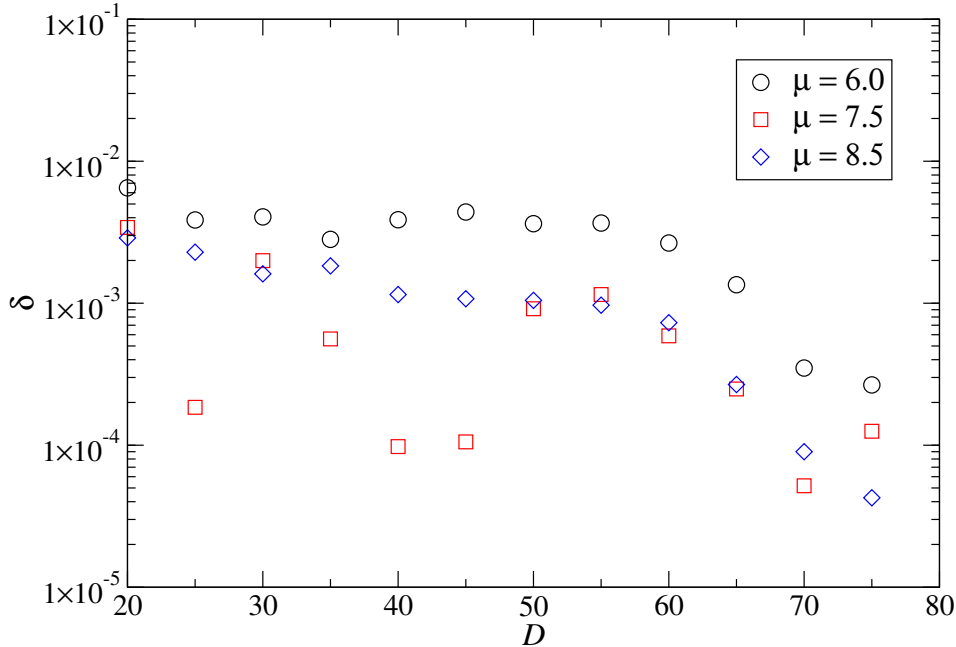}
	\caption{Convergence behavior of thermodynamic potential with $\delta$ of Eq.~\eqref{eq:delta} at $U=8$ with $\mu=6.0, 7.5$ and 8.5 as a function of $D$ on $V=4096^2\times 1677.7216$ lattice.}
  	\label{fig:lnZ_D}
\end{figure}

We investigate the convergence behavior of the thermodynamic potential by defining the quantity
\begin{align}
	\delta=\left|\frac{\ln Z(D)-\ln Z(D=80)}{\ln Z(D=80)}\right|
\label{eq:delta}
\end{align}
on $V=4096^2\times 1677.7216$ lattice with $\epsilon=10^{-4}$. In Fig.~\ref{fig:lnZ_D}, we plot the $D$ dependence of $\delta$  at $(U,t)=(8,1)$ with the choices of $\mu=6.0, 7.5$ and 8.5. As we will see below, $\mu=6.0$ corresponds to $\langle n\rangle\approx1.0$ and $\mu=8.5$ does to $\langle n\rangle\approx1.5$. We observe that $\delta$'s at these values of $\mu$ decrease as a function of $D$, though some of them are fluctuating.

\subsection{Strong coupling limit}

\begin{figure}
	\centering
	\includegraphics[keepaspectratio,scale=0.5]{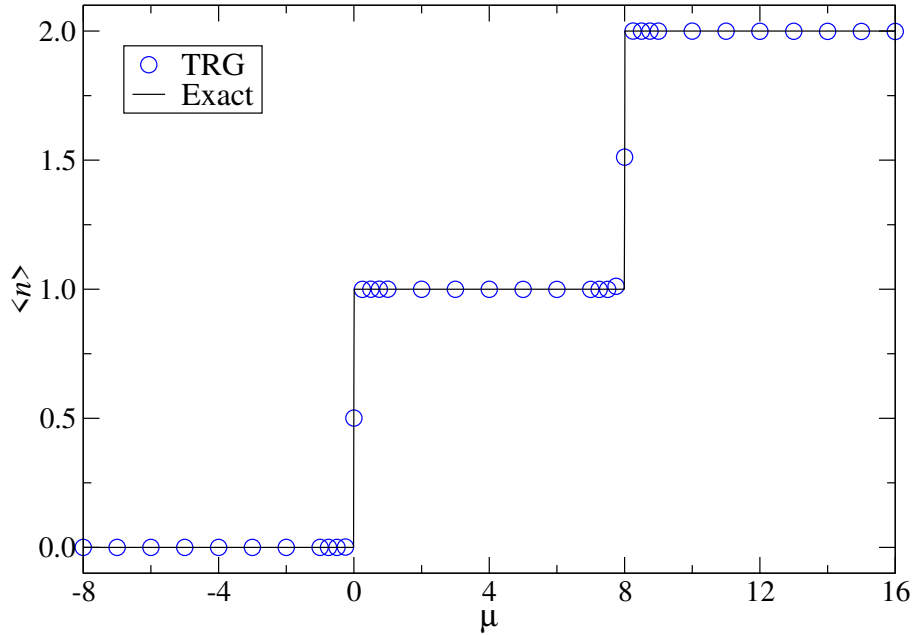}
	\caption{Electron density $\langle n\rangle$ in the $(U,t)=(8,0)$ case at $\beta=1677.7216$ with $\epsilon=10^{-4}$ as a function of $\mu$. The solid line shows the exact solution and the blue circles are the results obtained by the TRG.}
  	\label{fig:edensity_t0}
\end{figure}

We first consider the atomic limit at $(U,t)=(8,0)$. This case is analytically solvable. The electron density $\langle n\rangle$ is obtained by the numerical derivative of the thermodynamic potential in terms of $\mu$:
\begin{align}
	\langle n\rangle=\frac{1}{V}\frac{\partial \ln Z(\mu)}{\partial \mu}\approx
	\frac{1}{V}\frac{\ln Z(\mu+\Delta \mu)-\ln Z(\mu-\Delta \mu)}{2\Delta \mu}.
\end{align}
In Fig.~\ref{fig:edensity_t0} we compare the numerical and exact results for the $\mu$ dependence of $\langle n\rangle$ at $(U,t)=(8,0)$. Note that we set $m_{\tau}=24$ because this case is equivalent to the model defined on $V=1\times\beta$ lattice. Thanks to the vanishing hopping structure in the spatial direction, we can check the validity of the imaginary time evolution explained in Sec.~\ref{subsec:algorithm}.
The agreement of our numerical result with the exact solution shows that the imaginary time evolution carried out by the GATRG works precisely. 
The deviation from the exact value defined by
\begin{align}
	\delta_{\rm exact}(D)=\left|\frac{\ln Z(D)-\ln Z_{\rm exact}}{\ln Z_{\rm exact}}\right|
\label{eq:delta_exact}
\end{align}
is at most $O(10^{-4})$ in the range of $0\le \mu\le 16$. For $\mu<0$, the exact thermodynamic potential is equal to zero and the results obtained by the TRG are also equal to zero within a level of double precision.

\begin{figure}
	\centering
	\includegraphics[keepaspectratio,scale=0.5]{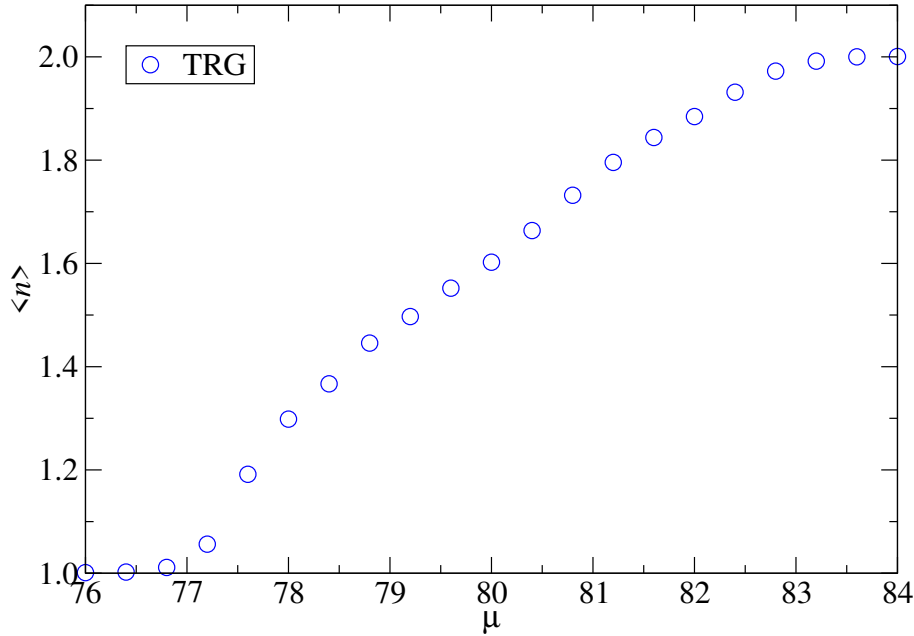}
	\caption{Electron density $\langle n\rangle$ as a function of $\mu$ at $(U,t)=(80,1)$ on $V=4096^2\times 1677.7216$ with $\epsilon=10^{-4}$. The bond dimension is set to be $D=80$.}
  	\label{fig:edensity_U80}
\end{figure}

Since the phase diagram of the metal-insulator transition in the (2+1)$d$ Hubbard model is not well known so far, we investigate the $\mu$ dependence of $\langle n\rangle$ choosing $U=80$ as a representative case in the strong coupling region. In Fig.~\ref{fig:edensity_U80} we plot the electron density  $\langle n\rangle$ as a function of $\mu$ in the vicinity of $\mu\sim U$ with $D=80$. We have checked that the convergence behavior of $\delta$ at $U=80$ is better than that at $U=8$.
We observe that the electron density starts to increase from $\langle n\rangle=1$ at $\mu=77.0(2)$ or $\mu/U=0.9625(25)$ and reaches $\langle n\rangle=2$ with $\mu\gtrsim83.0$ or $\mu/U\gtrsim 1.04$. The $\mu$ dependence of $\langle n\rangle$ is smooth and continuous so that there is no signal of the first-order phase transition.
We expect that the critical chemical potential $\mu_{\rm c}$ at the doping-driven metal-insulator transition approaches to $\mu_{\rm c}/U=1$ toward the atomic limit and the transition from $\langle n\rangle=1$ to $\langle n\rangle=2$ becomes a step-function as a function of $\mu/U$. 
 
\subsection{Critical chemical potential at $U=8$ and $2$}
\label{subsec:mu_c}

\begin{figure}
  	\centering
	\includegraphics[keepaspectratio,scale=0.5]{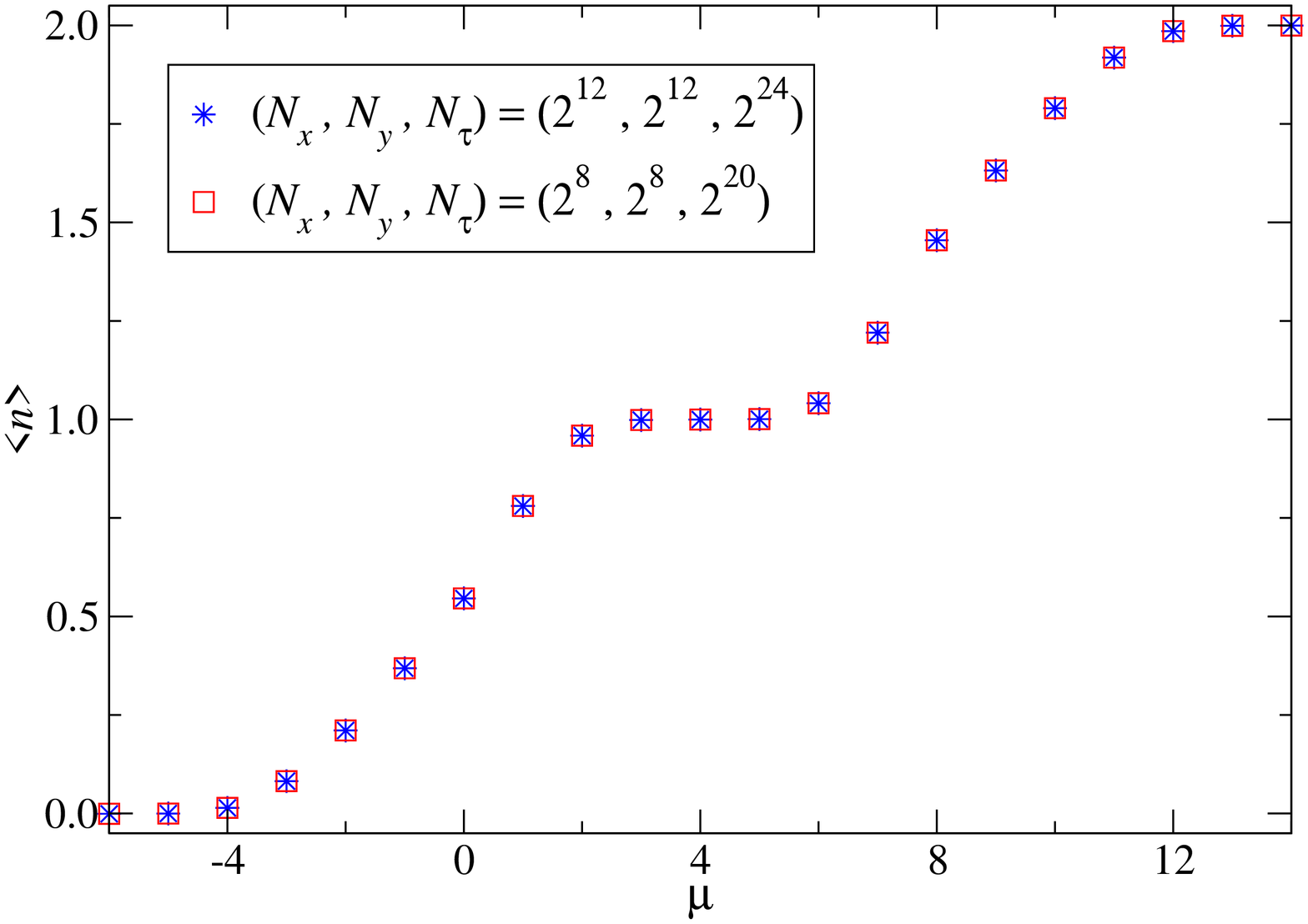}
	\caption{Electron density $\langle n\rangle$ at $U=8$ on two lattice sizes, $V=256^2\times 104.8576$ and $4096^2\times 1677.7216$, as a function of $\mu$. The bond dimension is set to be $D=80$.}
  	\label{fig:n_vol}
\end{figure}

Now let us investigate the metal-insulator transition in the intermediate coupling region at $(U,t)=(8,1)$. There are a lot of previous work to investigate a possible superconducting phase expected in this coupling region.  
Since we are interested in the thermodynamic and zero-temperature limit, we first check the volume dependence of the electron density $\langle n \rangle$.
In Fig.~\ref{fig:n_vol} we plot the $\mu$ dependence of $\langle n \rangle$ at $U=8$ changing the lattice sizes with $\epsilon=10^{-4}$, $m_{\tau}=12$ and $D=80$.  We observe that the size of $(N_{x},N_{y},N_{\tau})=(2^{12},2^{12},2^{24})$, which corresponds to $V=4096^2\times 1677.7216$, is sufficiently large to be identified as the thermodynamic and zero-temperature limit. We observe the $\langle n \rangle=0$ plateau for $\mu\lesssim-4$ and the $\langle n\rangle=2$ one for $12\lesssim \mu$. The half-filling state is characterized by the plateau of $\langle n\rangle=1$ in the range of $2\lesssim \mu\lesssim 6$. These plateaus yield the vanishing compressibility $\kappa=\partial\langle n\rangle/\partial \mu$ indicating the insulating states.

\begin{figure}
  	\centering
	\includegraphics[keepaspectratio,scale=0.5]{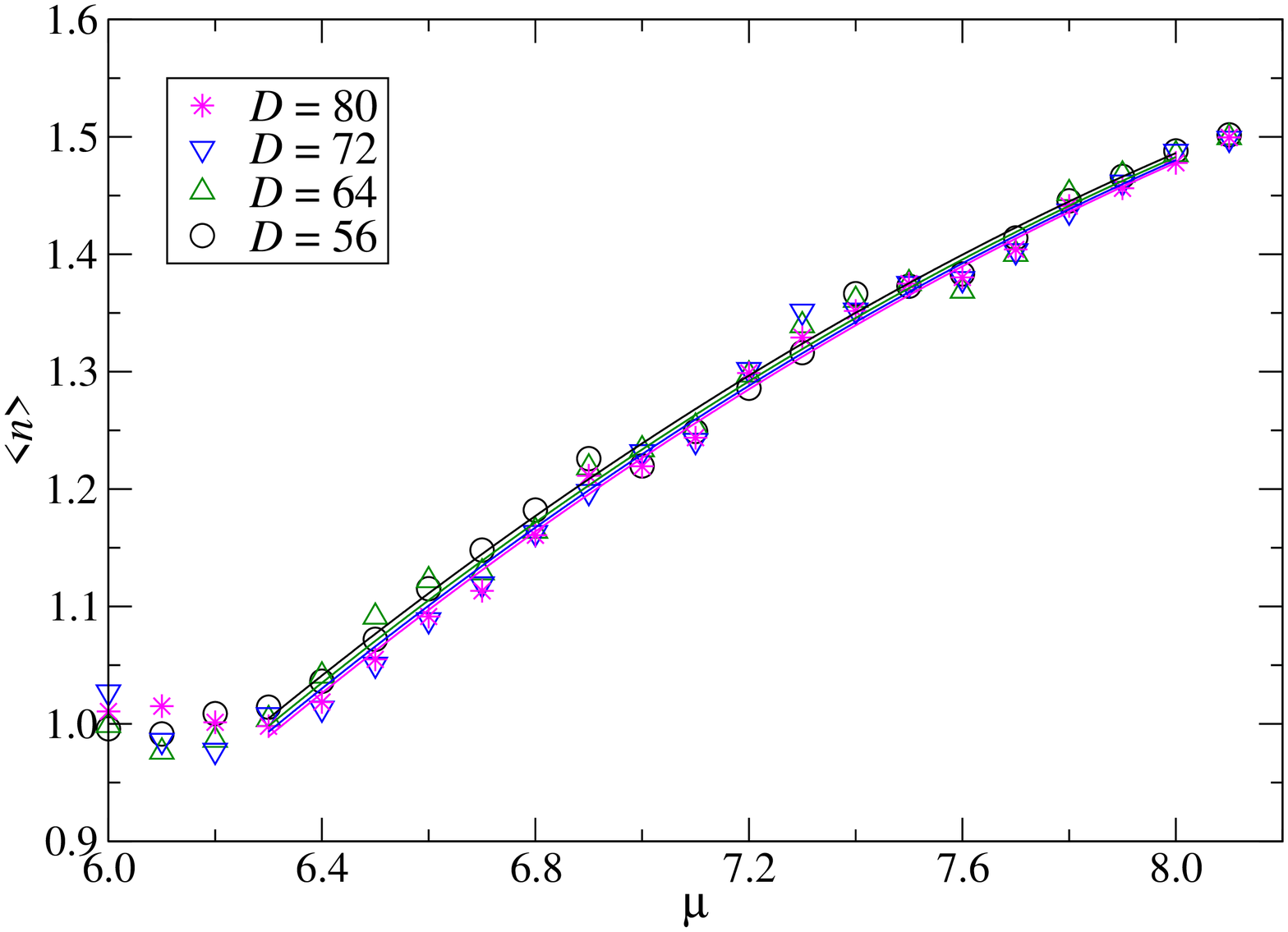}
	\caption{Electron density $\langle n\rangle$ at $U=8$ on $V=4096^2\times 1677.7216$  with $\epsilon=10^{-4}$ as a function of $\mu$. The bond dimensions are $D=80$, $72$, $64$ and $56$. Fit results are drawn by the solid lines for each bond dimension.}
  	\label{fig:n_u8}
\end{figure}

In order to determine the critical chemical potential $\mu_{\rm c}$ in the limit of $D\rightarrow\infty$ at $U=8$ on $V=4096^2\times 1677.7216$ lattice we make a global fit of $\langle n\rangle$ with $D=80$, 72, 64 and 56 in the metallic phase near the transition point. In Fig.~\ref{fig:n_u8} we plot the results of $\langle n\rangle$ at $D=80$, $72$, $64$ and $56$ with a much finer resolution of $\Delta \mu$ than Fig.~\ref{fig:n_vol} focusing on the range of $6.0\le \mu\le8.2$, which covers the region of $1\le\langle n\rangle\le1.5$. This figure provides us a closer look at the $\mu$ dependence of $\langle n\rangle$ around $\mu_{\rm c}$.  The results at $D=80$, $72$, $64$ and $56$ are almost degenerate indicating the small $D$ dependence. For the global fit we employ the following quadratic fitting function:
\begin{align}
	\langle n\rangle=1+\alpha\left(\mu-\mu_{\rm c}(D)\right)+\beta\left(\mu-\mu_{\rm c}(D)\right)^2
\label{eq:n_fit}
\end{align}
with $\mu_{\rm c}(D)=\mu_{\rm c}(D=\infty)+\gamma/D$, where $\alpha$, $\beta$, $\gamma$ and $\mu_{\rm c}(D=\infty)$ are the fit parameters. The solid curves in Fig.~\ref{fig:n_u8} represent the fit results over the range of $6.3\le \mu\le8.0$. We obtain $\mu_{\rm c}(D=\infty)=6.43(4)$, which is presented in Table~\ref{tab:mu_c} together with other fitting results. 
It may be instructive to compare Figs.~\ref{fig:n_vol}, \ref{fig:n_u8}, and the estimated location of $\mu_{\rm c}(D=\infty)$ with numerical data in Refs.~\cite{PhysRevB.67.085103,PhysRevB.87.035110}, though their calculations are carried out on very small lattice sizes and at low but finite temperatures.

\begin{table}[htb]
	\caption{Critical chemical potential $\mu_{\rm c}(D=\infty)$ at $U=8$ and $2$ determined by the global fit of the data at $D=80$, $72$, $64$ and $56$.}
	\label{tab:mu_c}
	\begin{center}
		\begin{tabular}{|c|cc|}\hline
    		$U$ & 8 & 2 \\ 
    		{\rm fit\; range} & [6.3, 8.0] & [1.2, 3.4] \\ \hline
    		$\mu_{\rm c}(D=\infty)$ & 6.43(4) & 1.30(6) \\ \hline
    		$\alpha$ & 0.372(9) & 0.39(1)  \\ \hline
    		$\beta$ & $-$0.051(6) & $-$0.054(5)  \\ \hline
    		$\gamma$ & $-$7(2) & $-$13(4)  \\ 
\hline
	\end{tabular}
	\end{center}
\end{table}

\begin{figure}
  	\centering
	\includegraphics[keepaspectratio,scale=0.5]{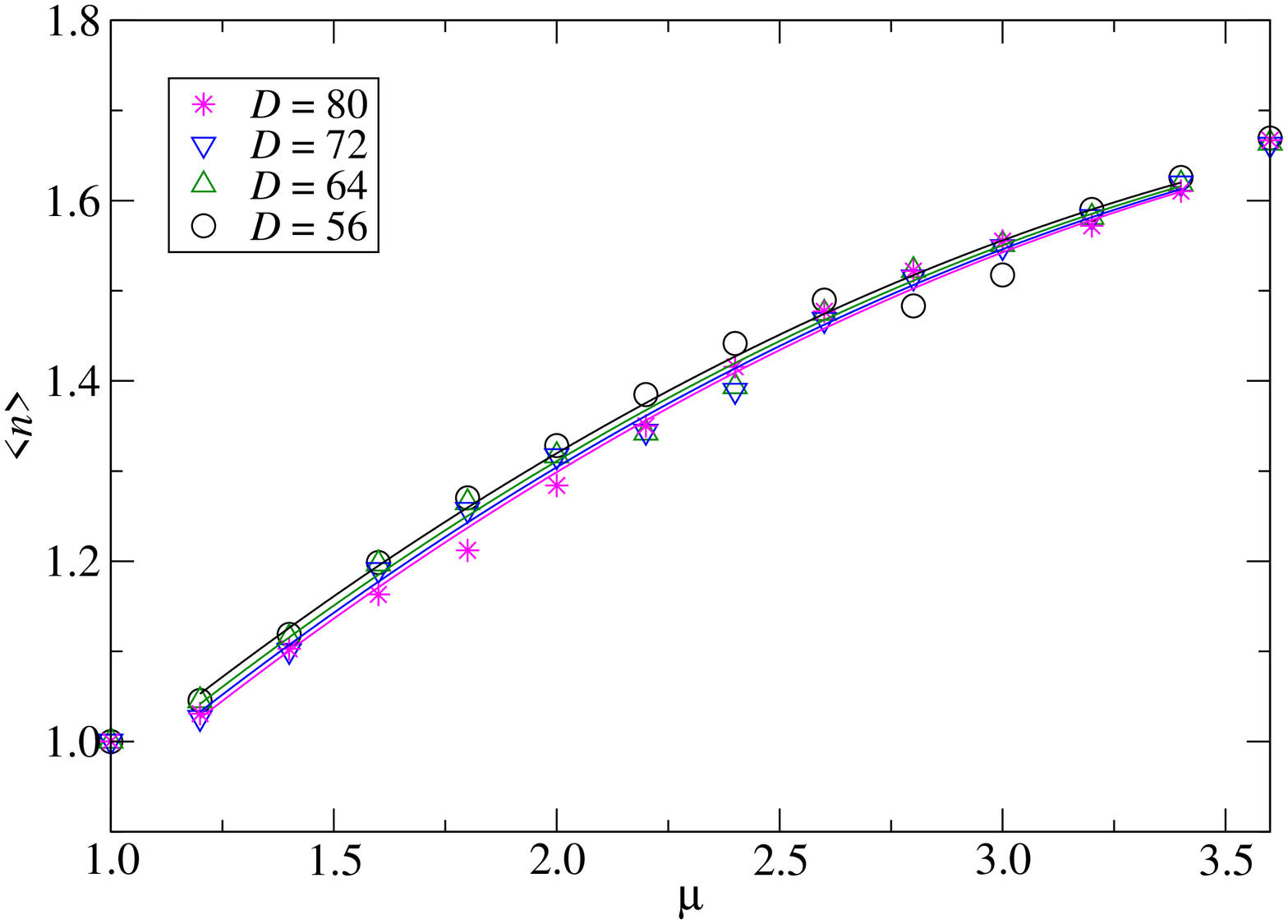}
	\caption{Electron density $\langle n\rangle$ at $U=2$ on $V=4096^2\times 1677.7216$  with $\epsilon=10^{-4}$ as a function of $\mu$. The bond dimensions are $D=80$, $72$, $64$ and $56$. Fit results are drawn by the solid lines for each bond dimension.}
  	\label{fig:n_u2}
\end{figure}

We repeat the same analysis for the weak coupling case at $U=2$.
We apply the fit function of Eq.~\eqref{eq:n_fit} to four data sets with the bond dimensions of $D=80$, 72, 64 and 56.
Fit results are depicted in Fig.~\ref{fig:n_u2} and their numerical values are summarized in Table~\ref{tab:mu_c}.
Our results show that the deviation of $\vert\mu_{\rm c}(D)-U/2\vert$ diminishes as the Coulomb potential $U$ decreases. It is likely that $\vert\mu_{\rm c}(D)-U/2\vert$ vanishes only at $U=0$.
This means that the model exhibits the metal-insulator transition over the wide regime of the finite coupling, including the weak coupling region.
This conclusion may provide us a different scenario of the phase diagram from that predicted by the dynamical mean-field theory (DMFT)~\cite{RevModPhys.68.13}; there exists some $U_{\rm c}$ such that no metal-insulator transition occurs with $U<U_{\rm c}$.

\section{Summary and outlook} 
\label{sec:summary}

We have investigated the doping-driven metal-insulator transition of the (2+1)$d$ Hubbard model in the path-integral formalism employing the TRG method. The electron density $\langle n\rangle$ is calculated in the wide range of $\mu$ corresponding to $0\le\langle n\rangle\le 2$.
We have also determined the critical chemical potential $\mu_{\rm c}$ at three values of $U$. Our results indicate that the deviation $\vert \mu_{\rm c}-U/2\vert$ vanishes only at $U=0$. 
This means that the model exhibits the metal-insulator transition over the vast regime of the finite coupling $U$.
As a next step, it would be interesting to investigate the metal-insulator transition of the (3+1)$d$ Hubbard model.


\section*{Acknowledgment}

Numerical calculation for the present work was carried out with the supercomputer Fugaku provided by RIKEN (Project ID: hp200314) and also with the Oakforest-PACS (OFP) computer under the Interdisciplinary Computational Science Program of Center for Computational Sciences, University of Tsukuba. This work is supported in part by Grants-in-Aid for Scientific Research from the Ministry of Education, Culture, Sports, Science and Technology (MEXT) (No. 20H00148) and JSPS KAKENHI Grant Number JP21J11226 (S.A.).


\bibliographystyle{ptephy}
\bibliography{bib/formulation,bib/algorithm,bib/discrete,bib/grassmann,bib/continuous,bib/gauge,bib/review,bib/for_this_paper}
%





\end{document}